\documentclass{cernrep}
\usepackage{graphicx,here}
 
\begin{document}

\title{Uncertainties in Parton Related Quantities}
 
\author{R.S. Thorne}
 
\institute{Cavendish Laboratory, University of Cambridge, UK}
%
%
%
 
\def\GeV{{\rm GeV}}
 
\maketitle 
 
\begin{abstract}
I discuss the issue of uncertainties in parton distributions and in the 
physical quantities which are determined in terms of them. While there has 
been significant progress on the uncertainties associated with errors on
experimental data, there are still outstanding questions. Also, I demonstrate
that in many circumstances this source of errors may be less important than 
errors due to underlying assumptions in the fitting procedure and due to the 
incomplete nature of the theoretical calculations. 
\end{abstract}
 
\section{Introduction to Global Fits}
 
The fundamental quantities 
one requires in the calculation of scattering 
processes involving hadronic particles are  the parton distributions. 
These can be derived from and then used within QCD. 
Using the Factorization Theorem 
the cross-section for this process 
can be written in the factorized form
\begin{equation}
\sigma(ep \to eX) = \sum_i C^P_{i}(x,\alpha_s(Q^2))\otimes
f_{i}(x,Q^2,\alpha_s(Q^2)) 
\label{a}
\end{equation}
up to corrections of order $\Lambda_{{\rm QCD}}^2/Q^2$, 
known as higher twist. 
The coefficient functions $C^P_{i}(x,\alpha_s(Q^2))$ 
describing the hard 
scattering process are process dependent but are calculable as a power-series
in the strong coupling constant $\alpha_s(Q^2)$.
\begin{equation}
C^P_{i}(x,\alpha_s(Q^2))= \sum_k C^{P,k}_{i}(x)\alpha^k_s(Q^2).
\label{c}
\end{equation}
The $f_{i}(x,Q^2,\alpha_s(Q^2))$ 
are the parton distributions, i.e the 
probability of finding a parton of type $i$ carrying a fraction $x$ 
of the momentum of the hadron. 
Because they depend on the nonperturbative way 
in which partons are bound into the hadron, these parton distributions are 
not calculable from first principles. However, they do evolve with $Q^2$ 
in a perturbative manner
\begin{equation}
\frac{d f_{i}(x,Q^2,\alpha_s(Q^2))}{d \ln Q^2} = \sum_i 
P_{ij}(x,\alpha_s(Q^2))\otimes f_{j}(x,Q^2,\alpha_s(Q^2)) 
\label{b}
\end{equation}
where the splitting functions $P_{ij}(x,Q^2,\alpha_s(Q^2))$
are calculable order by order in 
perturbation theory. 
Since the parton distributions $f_{i}(x,Q^2,\alpha_s(Q^2))$ 
are process-independent, i.e. universal, once they have been measured at 
one experiment, one can predict many other scattering processes. 

In order to determine the parton distributions 
one can use a range of available data -- 
largely $ep \to eX$ (structure functions), and the most 
up-to-date QCD calculations, which are currently NLO-in-$\alpha_s(Q^2)$.
(NNLO coefficient functions are known for some processes, e.g. 
structure functions, and NNLO splitting functions have considerable 
information, and may be known within a year or so.) 
Perturbation theory is assumed to be valid if 
$\alpha_s(Q^2)< 0.3$ so only data with $Q^2 > 2 {\rm GeV^2}$ or 
more are used. This cut should also remove the influence of higher twists.

The global fit \cite{GRV}-\cite{ref:giele}  
usually proceeds by starting the parton evolution at a low 
scale $Q_0^2 \sim 1 \GeV^2$, and evolving partons upwards using NLO
DGLAP equations. In principle there are 11 different parton 
distributions to consider (Isospin symmetry is assumed, i.e. if $p \to n$, 
$d(x) \to u(x)$ and $u(x) \to d(x)$.)
\begin{equation}
u, \bar u, \quad d, \bar d, \quad s, \bar s, \quad c, \bar c, 
\quad b, \bar b, \quad g. 
\label{f}
\end{equation}
In practice $m_c, m_b \gg \Lambda_{{\rm QCD}}$ so the heavy 
parton distributions are
determined perturbatively. Also it is 
currently assumed that $s=\bar s$. 
The 6 independent parton sets are then
\begin{equation}
u_V = u- \bar u, \quad d_V =d-\bar d, 
\quad {\rm sea}=2*(\bar u + \bar d + 
\bar s), \quad \bar d - \bar u, \quad g.
\label{g}
\end{equation}

The input partons are parameterized in a particular form, e.g.
\begin{equation}
xf(x, Q_0^2) = A(1-x)^{\eta}(1+\epsilon x^{0.5}+\gamma x)
x^{\delta}.
\end{equation}  
The partons are then constrained by a number of sum rules:
\begin{equation}
\int_0^1 u_V(x)\, dx =2 \qquad \int_0^1 d_V(x)\, dx =1 \qquad 
\int_0^1 x\Sigma(x) +x g(x) \, dx =1,
\label{i}
\end{equation}
i.e. conservation of the number of valence quarks, and
conservation of the momentum carried by partons.
The latter is an important constraint on the form of the gluon which is only 
probed indirectly. 

In determining partons one needs to consider that not only are there 
6 different 
combinations of partons, but there is also a wide distribution of
$x$ from $0.75$ to $0.00003$. 
One needs many different types of experiment for
full determination.   
The full set of data usually used is 
H1 and ZEUS $F^p_2(x,Q^2)$ data \cite{H1A,ZEUS} which covers 
small $x$ and a wide range of $Q^2$; E665 
$F^{p,d}_2(x,Q^2)$ data \cite{E665} at medium $x$;
BCDMS and SLAC $F^{p,d}_2(x,Q^2)$ data \cite{BCDMS}-\cite{SLAC} at large $x$; 
NMC $F^{p,d}_2(x,Q^2)$ \cite{NMC} at medium and large $x$; 
CCFR $F^{\nu(\bar\nu) p}_2(x,Q^2)$ and
$F^{\nu(\bar\nu) p}_3(x,Q^2)$ data \cite{CCFR} at large $x$ 
which probe the singlet and valence quarks independently;
ZEUS  and H1 $F^{p}_{2,charm}(x,Q^2)$ data \cite{ZEUSc,H1c}; 
E605 $ p N \to \mu \bar \mu + X$ \cite{E605} constraining the large $x$ 
sea; E866 Drell-Yan asymmetry \cite{E866} which determines  
$\bar d -\bar u$; CDF W-asymmetry data \cite{Wasymm} 
which constrains the $u/d$ ratio at 
large $x$; CDF and D0 inclusive jet data \cite{D0,CDF} 
which tie down the high $x$ gluon; 
and CCFR and NuTev Dimuon data \cite{CCFRdimuon, NuTeV}
which constrain the strange sea.
Note that I discuss unpolarized parton distributions. 
There are far fewer data for polarized distributions,
though fits with error determinations do exist, e.g. \cite{spin}.

\subsection{Quality of the Fit}

This is determined by the $\chi^2$ of the fit to 
data, which may be calculated in various ways. 
The simplest is to add statistical and systematic errors in quadrature. 
This ignores correlations between data points, but is sometimes 
quite effective. Also, the information on the data often 
means that only this method is available.
 
However, more properly one uses the full covariance matrix which 
is constructed as
\begin{equation}
C_{ij} = \delta_{ij} \sigma_{i,stat}^2 + \sum_{k=1}^n \rho^k_{ij}
\sigma_{k,i}\sigma_{k,j},
\label{k}
\end{equation}
where $k$ runs over each source of correlated systematic error
and $\rho^k_{ij}$ are the correlation coefficients. The $\chi^2$ 
is defined by
\begin{equation}
\chi^2 = \sum_{i=1}^N\sum_{j=1}^N (D_i-T_i(a))C^{-1}_{ij}
(D_j-T_j(a)),
\label{l}
\end{equation}
where $N$ is the number of data points, $D_i$ is the 
measurement
and $T_i(a)$ is the theoretical prediction depending on parton input 
parameters $a$. Unfortunately this method relies on inverting very 
large matrices. 

An alternative which is identical to the correlation matrix 
definition of $\chi^2$ if the errors are small is to 
incorporate the correlated errors into the theory prediction
\begin{equation}
f_i(a,s) = T_i(a) + \sum_{k=1}^n s_k \Delta_{ik},
\label{m}
\end{equation}
where $\Delta_{ik}$ is the one-sigma correlated error for point 
$i$ from source $k$. In this case the $\chi^2$ is defined by     
\begin{equation}
\chi^2 = \sum_{i=1}^N \biggl(\frac{D_i-f_i(a,s)}{\sigma_{i,unc}}
\biggr)^2 + \sum_{k=1}^n s_k^2,
\label{n}
\end{equation}
where the second term constrains the values of $s_k$,
assuming the correlated systematic errors are Gaussian distributed. 
In this method
the data may move {\it en masse} relative to the theory. 
One can solve for the $s_k$ analytically \cite{ref:lmethod,CTEQ6}. 
Defining
\begin{equation}
B_k = \sum_{i=1}^N \frac{\Delta_{ik}(D_i-T_i(a))}{\sigma_{i,unc}^2}, 
\qquad
A_{kl}= \delta_{kl} + \sum_{i=1}^N \frac{\Delta_{ik}\Delta_{il}}
{\sigma_{i,unc}^2}
\label{o}
\end{equation}
one obtains
\begin{equation}
\frac{\partial \chi^2}{\partial s_k} =0 \quad \to \quad
s_i(a) = \sum_{l=1}^n (A^{-1})_{kl}B_l.
\label{p}
\end{equation}
This leads to the $\chi^2$ definition 
\begin{equation}
\chi^2 = \sum_{i=1}^N \biggl(\frac{(D_i-T_i(a))}{\sigma_{i,unc}}
\biggr)^2 - \sum_{k=1}^n\sum_{l=1}^n B_k (A^{-1})_{kl} B_l.
\label{q}
\end{equation}
This approach has the double advantage that smaller matrices need 
inverting and one sees explicitly the shift of data relative to theory.  
However, it is doubtful that Gaussian correlated errors 
are realistic. The method also allows one to 
move data simply to compensate for the shortcomings of theory. Indeed, 
MRST find that for HERA data increments in $\chi^2$ 
using this method are the same as for adding in quadrature, and that the 
data move towards theory rather than {\it vice versa} \cite{MRST01}.
Hence it is questionable in practice quite how much of an improvement this 
approach is in many cases. However, for Tevatron jet data, 
where correlated systematic errors dominate,
a sophisticated treatment of correlated errors is essential.

Using some particular method of calculating $\chi^2$ the global fit 
procedure completely determines parton distributions at present. 
In general the total fit is of reasonably good quality, 
as illustrated for the major data sets, and the CTEQ6 fit 
(which assumes $ \alpha_S(M_Z^2)$ fixed at $0.118$) in table 1. 
The total $\chi^2=1954/1811$.
For MRST  $\alpha_S(M_Z^2)$ is determined to be $0.119$, and the  
total $\chi^2=2328/2097$.
However, the $\chi^2$ per point of more than one suggests some possible 
shortcomings, and it may be argued that there are some areas where the 
theory perhaps needs to be improved.

\vspace{-0.4cm}
\begin{center}
{\footnotesize A table of $\chi^2$ versus no. 
of data points for the CTEQ6 fit.}\\
\vspace{0.2cm}
\begin{tabular}{|ccc|}\hline                                                 
Data set          & No. of   & $\chi^2$ \\                   
                  & data pts &      \\ \hline                   
H1 $ep$           & 230      & 228   \\                   
ZEUS $ep$         & 229      & 263    \\                   
BCDMS $\mu p$     & 339      & 378    \\
BCDMS $\mu d$     & 251      & 280    \\                   
NMC $\mu p$       & 201      & 305   \\                                
E605 (Drell-Yan) & 119      &  95   \\        
D0 Jets           &  90      &  65  \\ 
CDF Jets          &  33      &  49  \\ \hline                   
\end{tabular}
\end{center}

\section{Parton Uncertainties}  

There are a number of different approaches for obtaining parton uncertainties. 

\subsection{Hessian (Error Matrix) Approach} 

This was first used by H1 and has recently been extended by CTEQ. 
One defines the Hessian matrix by
\begin{equation}
\chi^2 -\chi_{min}^2 \equiv \Delta \chi^2 = \sum_{i,j} 
H_{ij}(a_i -a_i^{(0)})
(a_j -a_j^{(0)})
\label{r}
\end{equation}
\noindent
The Hessian matrix $H$ is related to the covariance matrix 
of the parameters by
\begin{equation}
C_{ij}(a) = \Delta \chi^2 (H^{-1})_{ij}.
\label{s}
\end{equation}
We can then use the standard formula for linear error propagation:  
\begin{equation}
(\Delta F)^2 = \Delta \chi^2 \sum_{i,j} \frac{\partial F}
{\partial a_i}(H)^{-1}_{ij}  
\frac{\partial F}{\partial a_j}.
\label{t}
\end{equation}
This has been used to find partons with errors by $H1$ \cite{ref:h1fit}
and Alekhin \cite{ref:alekhinfit}, each with restricted data sets.
\begin{figure}[htbp]
\begin{center}
\includegraphics[width=12cm]{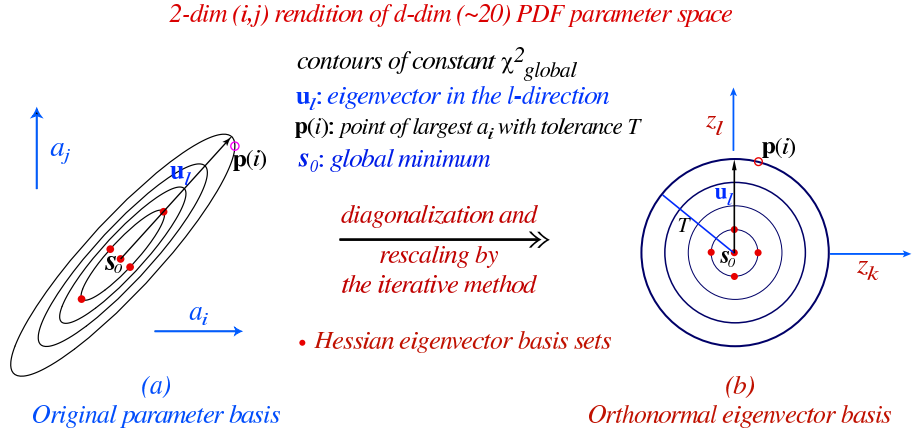}
\end{center}
\vspace{-0.8cm}
\caption{Representation of diagonalization of Hessian matrix.}
\label{one}
\end{figure}  
In practice it is problematic due to extreme 
variations in $\Delta \chi^2$ in different directions in parameter space.
This is solved by finding and rescaling eigenvectors of $H$ 
leading to the diagonal form 
\begin{equation}
\Delta \chi^2 = \sum_{i} z_i^2. 
\label{u}
\end{equation}
The method has been implemented by CTEQ \cite{ref:cteqpap1,ref:hmethod,CTEQ6}. 
The uncertainty on a physical quantity is 
\begin{equation}
(\Delta F)^2 = \sum_{i} \bigl(F(S_i^{(+)})-F(S_i^{(-)})\bigr)^2,
\label{v}
\end{equation}
where $S_i^{(+)}$ and $S_i^{(-)}$ are PDF sets 
displaced along eigenvector
directions by the given $\Delta \chi^2$. There is uncertainty in choosing 
the ``correct'' $\Delta \chi^2$ (in principle one unit) 
given the complications of a full global fit. CTEQ
choose $\Delta \chi^2 \sim 100$ \cite{ref:lmethod}. 
A discussion of this problem is found in \cite{statsa}. 

\begin{figure}[htbp]
\begin{center}
\includegraphics[width=7cm]{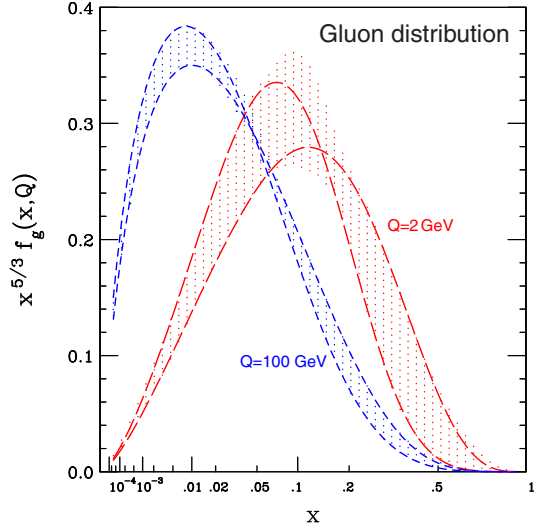}
\end{center}
\vspace{-0.8cm}
\caption{Results of CTEQ Hessian approach for gluon uncertainty.}
\label{two}
\end{figure}

\subsection{The Offset Method.}  

In this case the best fit is obtained by minimizing
\begin{equation}
\chi^2 = \sum_{i=1}^N \biggl(\frac{(D_i-f_i(a,s))}{\sigma_{i,unc}}
\biggr)^2,
\label{w}
\end{equation}
i.e. the best fit and parameters $a_0$ are obtained by considering only 
uncorrelated 
errors. This forces the theory to be close to unshifted data. 
The quality of the fit is then 
estimated by adding errors in quadrature. The systematic errors on the 
$a_i$ are determined by letting each $s_k = \pm 1$ and adding 
the deviations in 
quadrature. In practice one calculates 2 Hessian matrices
\begin{equation}
M_{ij} = \frac {\partial ^2 \chi^2}{\partial a_i \partial a_j}
\qquad  V_{ij} = \frac {\partial ^2 \chi^2}{\partial a_i \partial s_j},
\label{x}
\end{equation}
and defines covariance matrices  
\begin{equation}
C_{stat} = M^{-1} \qquad C_{sys} = M^{-1}VV^TM^{-1} \qquad C_{tot} = 
C_{stat} + C_{sys}.
\label{y}
\end{equation}
to achieve the same result.
This was used in early H1 fits \cite{ref:pascaud} and by 
ZEUS. A discussion and presentation of this method and of 
ZEUS results can be found in 
\cite{Mandy}. The offset method leads to a bigger uncertainty than the 
Hessian method for the same 
$\Delta \chi^2$ \cite{ref:alekhinstudy}. 

\subsection{Statistical Approach\cite{ref:giele}} 

\begin{figure}[htbp]
\vspace{-1cm}
\begin{center}
\includegraphics[width=10.0cm]{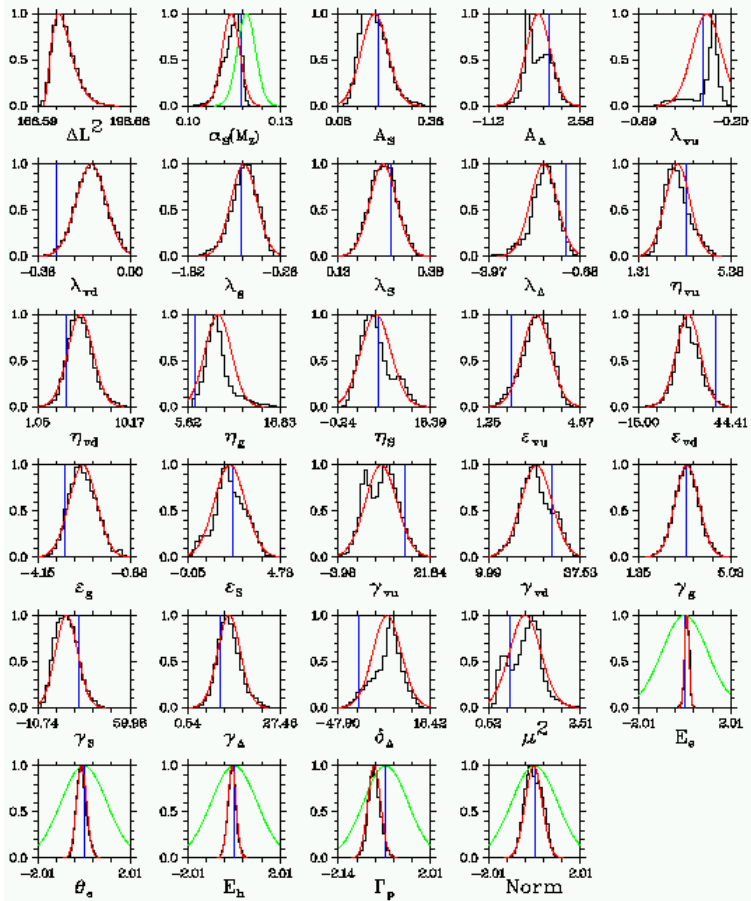}
\end{center}
\vspace{-1cm}
\caption{One set of parton parameters obtained from \cite{ref:giele}. 
The red curve is the Gaussian approximation
and the blue line the MRST value. The green curve for $\alpha_S$ is the 
LEP result.}
\label{three}
\end{figure}

In this one constructs an ensemble of distributions 
labelled by ${\cal F}$ each with 
probability $P(\{{\cal F}\})$. The mean $\mu_O$ and deviation 
$\sigma_O$ 
of observable $O$ are then 
given by
\begin{equation}
\mu_O = \sum_{\{{\cal F}\}}O(\{{\cal F}\})P(\{{\cal F}\}), \quad
\sigma_O^2 = \sum_{\{{\cal F}\}}(O(\{{\cal F}\})-\mu_O)^2 P(\{{\cal F}\}).
\label{aa}
\end{equation}
While this is statistically correct, and does not rely on 
the approximation of linear propagation of errors in calculating observables,
it is inefficient. In practice, 
one generates $N_{pdf}$ different distributions
with unit weight but distributed according to $P(\{{\cal F}\})$ where 
$N_{pdf}$ can be made as small as 100. Then 
\begin{equation}
\mu_O = \frac {1}{N_{pdf}}\sum_1^{N_{pdf}}O(\{{\cal F}\}), \quad
\sigma_O^2 =\frac {1}{N_{pdf}} \sum_{1}^{N_{pdf}}(O(\{{\cal F}\})-\mu_O)^2.
\label{ab}
\end{equation}
One can incorporate full information about measurements and their error 
correlations in the calculation of $P(\{{\cal F}\})$.

Currently the authors of \cite{ref:giele} use only proton 
DIS data sets in order to avoid complicated 
uncertainty issues such as shadowing effects for nuclear targets. 
Using strict confidence limits they find it difficult to obtain 
consistency between many different DIS experiments. Also the lack of important 
data sets leads 
to ``unusual'' values for some parameters, which illustrates the 
importance of using a wide variety of data.  
However, fig. 3 shows that indeed the Gaussian approximation is often not 
good, and shows potential complications for the more simplistic approaches.
This is a very attractive but ambitious large-scale project with a lot of 
work still to be done.  

\subsection{Lagrange Multiplier}

One can look at the uncertainty on a given physical quantity using the 
Lagrange Multiplier method, first suggested by CTEQ \cite{ref:lmethod} 
and also used by MRST \cite{MRSTBologna,MRSTL}. One performs 
the global fit while constraining the value of some physical 
quantity, i.e.  minimizing 
\begin{equation}
\Psi(\lambda,a) = \chi^2_{global}(a)  + \lambda F(a)
\label{ac}
\end{equation}
for various values of $\lambda$. This gives the set of best fits for 
particular 
values of the parameter $F(a)$ without relying on the Gaussian approximation
for $\Delta\chi^2$. A useful example is the $W$ cross-section at
Tevatron which is illustrated in fig. 4. 
The uncertainty in a quantity is determined by deciding an 
allowed value of $\Delta \chi^2$.

\begin{figure}[htbp]
\begin{center}
\includegraphics[width=5cm]{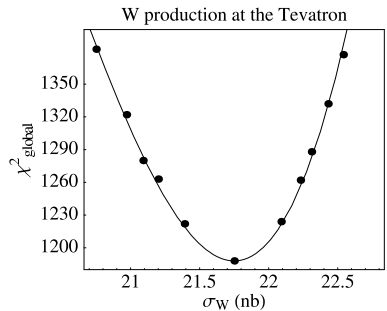}
\end{center}
\vspace{-0.8cm}
\caption{Variation of $\sigma_W$ with total $\chi^2$ for the CTEQ fit.}
\label{four}
\end{figure}
CTEQ use $\Delta \chi^2 =100$ (same as for the Hessian approach). 
They obtain for $\alpha_S =0.118$ \cite{CTEQ6} 
\begin{eqnarray}
\Delta \sigma_{W}(\rm LHC) &\approx& \pm 4\% \quad 
\Delta \sigma_{W}({\rm Tev}) \approx \pm 5\%\nonumber\\ 
\Delta 
\sigma_{H}({\rm LHC}) &\approx& \pm 5\%.
\label{ad}
\end{eqnarray}
The procedure is also used by MRST for a wider range of data, 
and using $\Delta \chi^2 \sim 50$. They find that 
for $\alpha_S =0.119$ \cite{MRSTL} 
\begin{eqnarray}
\Delta \sigma_{W}(\rm Tev) &\approx& \pm 1.2\% \quad 
\Delta \sigma_{W}({\rm LHC}) \approx \pm 2\%\nonumber\\
\Delta \sigma_{H}(\rm Tev) &\approx& \pm 4\% \quad 
\Delta \sigma_{H}({\rm LHC}) \approx \pm 2\%.
\label{ae}
\end{eqnarray}
If $\alpha_S$ also varies, $\Delta \sigma_{W}$ is quite 
stable but $\Delta \sigma_{H}$ almost doubles. The $\chi^2$ profile is 
shown in fig. 5.    
One can repeat for other processes, e.g. HERA charged current data are 
sensitive to very high $x$ quarks, the Tevatron jet data is sensitive to 
high $x$ gluon {\it etc.}.  

Overall one concludes that the uncertainty due to experimental errors 
is rather small, however they are dealt with.
It only exceeds a few $\%$ for quantities related to the high $x$ gluon
or very high $x$ quarks. However, there are other sources of error.

\vspace{-0.2cm}
\begin{figure}[htbp]
\includegraphics[width=6.7cm]{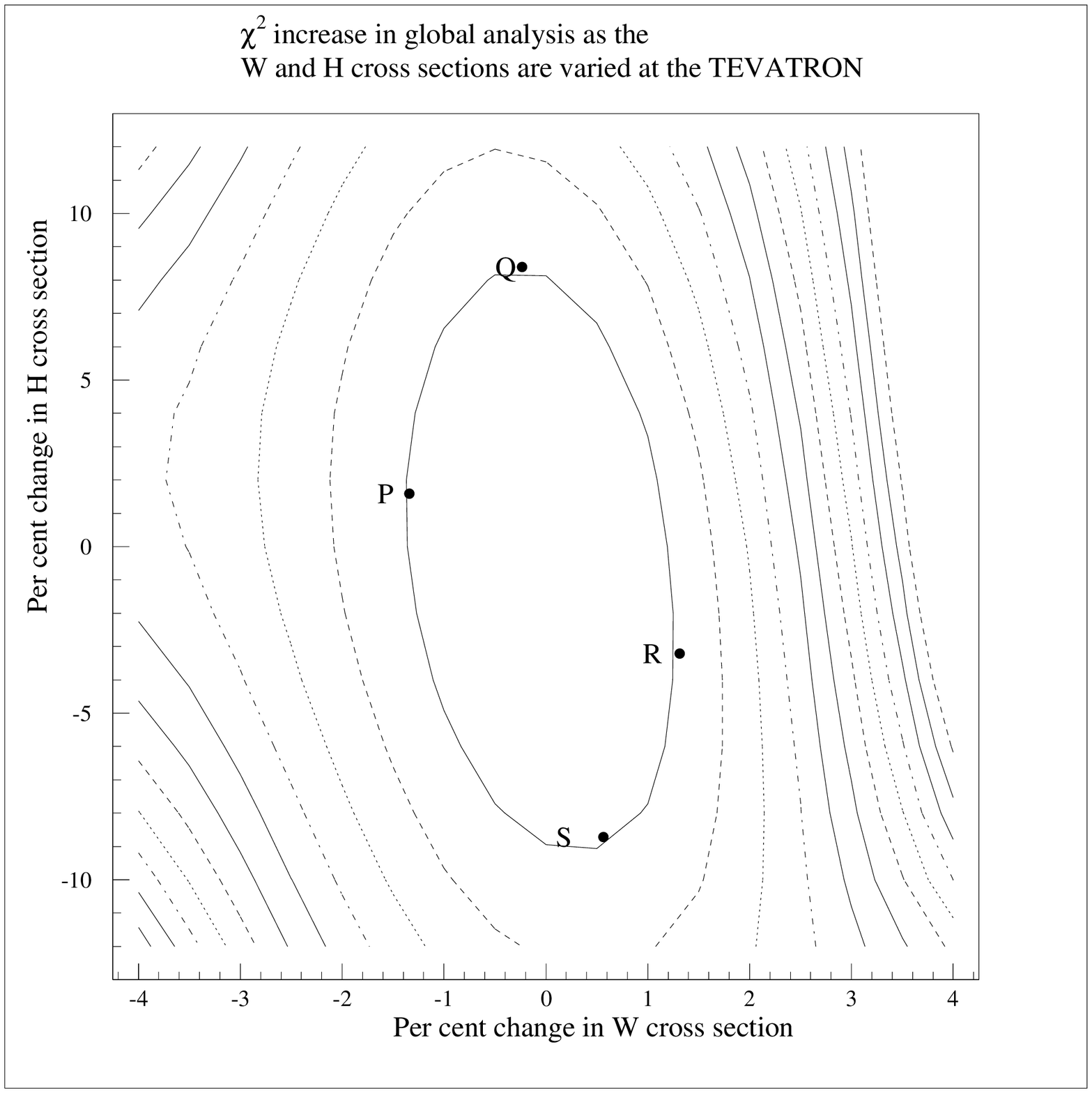}
\hspace{1.5cm}
\includegraphics[width=6.7cm]{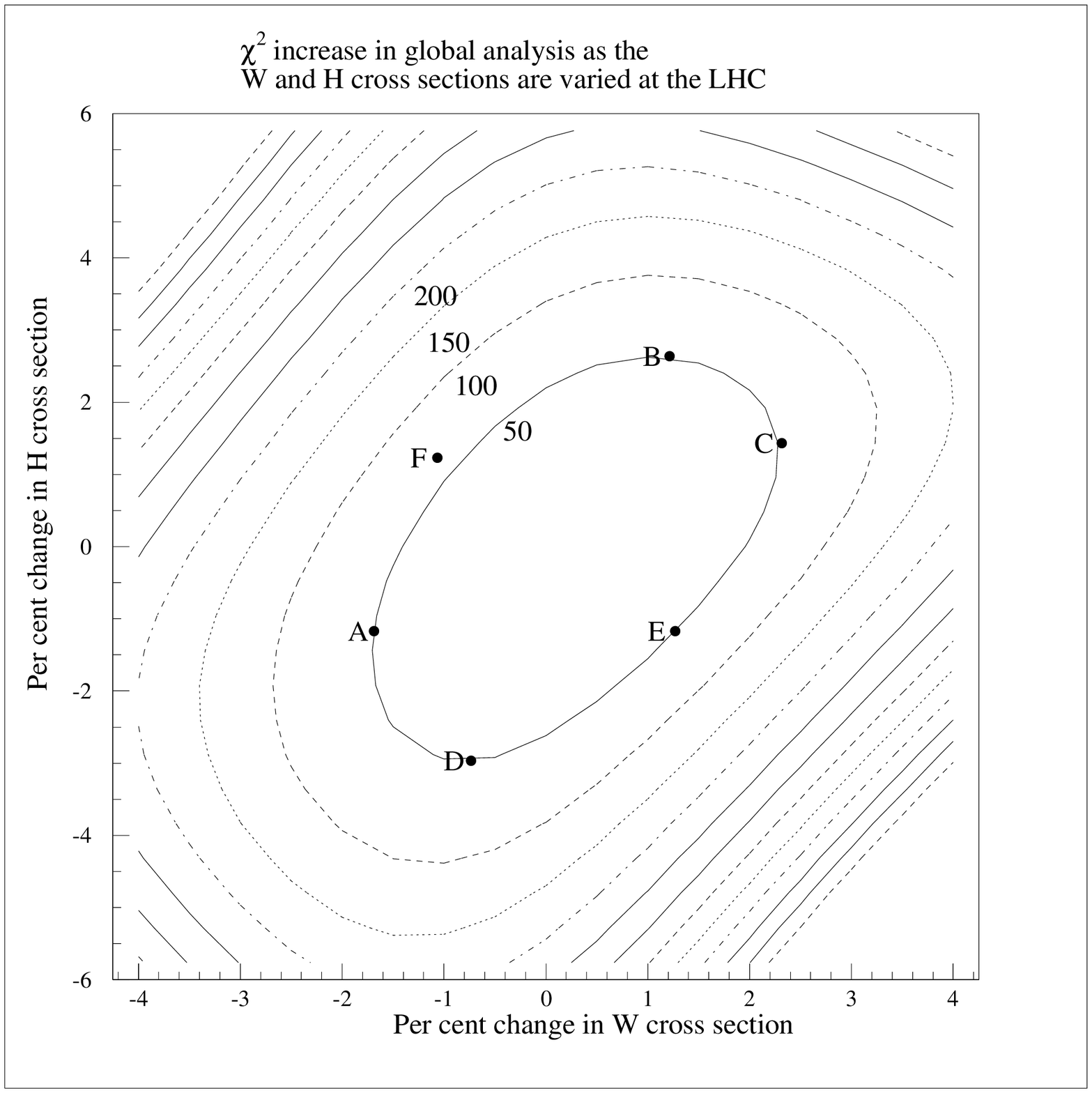}
\vspace{-2.2cm}
\caption{$\Delta\chi^2$-plot for $W$ and Higgs production at the 
Tevatron and LHC with $\alpha_S$ free. Contours show increments of 50 
in $\Delta \chi^2$.}
\label{five}
\end{figure}

\vspace{-0.4cm}

\section{Other Errors.}

To obtain a complete estimate of errors, one also 
needs to consider the effect of the assumptions made during 
the fit. These include the cuts made on the data, 
the data sets fit, the parameterization for the input sets, 
the form of strange sea, the assumption of no isospin violation, 
{\it etc}.. It is known that many of these can be as important as 
the experimental errors on data used (or even more so). A more 
systematic study is needed. 

It is also vital to consider sources of theoretical error. 
These include higher twist at low $Q^2$
and higher orders in $\alpha_S$. The latter are due not only to NNLO 
corrections, but also to enhancements at 
large and small $x$ because of terms of the form 
$\alpha_s^n \ln^{n-1}(1/x)$ and $\alpha_s^n 
\ln^{2n-1}(1-x)$ in the perturbative expansion.  
This means that renormalization and factorization 
scale variation are not a reliable way of estimating higher 
order effects, e.g., at small $x$
\begin{equation}
P^1_{qg} \sim \alpha_S(\mu^2) \qquad \qquad P^2_{qg} 
\sim \frac{\alpha_s(\mu^2)}{x}
\label{af}
\end{equation}
whereas
\begin{equation}
P^n_{qg} \sim \frac{\alpha_S^n(\mu^2) \ln^{n-2}(1/x)}{x}.
\label{ag}
\end{equation}
and scale variations of $P^1_{qg}, P^2_{qg}$ 
never give an indication of these terms. Hence,
in order to investigate the 
\newpage
\noindent true theoretical error we must consider 
some way of performing correct large and small $x$ resummations,
and/or use what we already know about NNLO. The latter approach 
implies that some quantities may acquire large higher order corrections
\cite{MRSTNNLO}.   

Alternatively, one can use the 
empirical approach of investigating in detail the effect of cuts on data.
In order to investigate the real quality of the fits and the regions with 
potential problems we try  
changing $W^2_{cut}$, $Q^2_{cut}$ and $x_{cut}$, 
re-fitting and seeing if the fit to the remaining data improves 
and/or the input parameters change dramatically \cite{MRSTcuts}. 
(Similar to a previous suggestion in terms of data 
sets rather than region of parameter space \cite{Collins}.) 
This is continued until the fit quality and the partons stabilize. 

\vspace{0.2cm}
\begin{figure}[htbp]
\includegraphics[width=7cm]{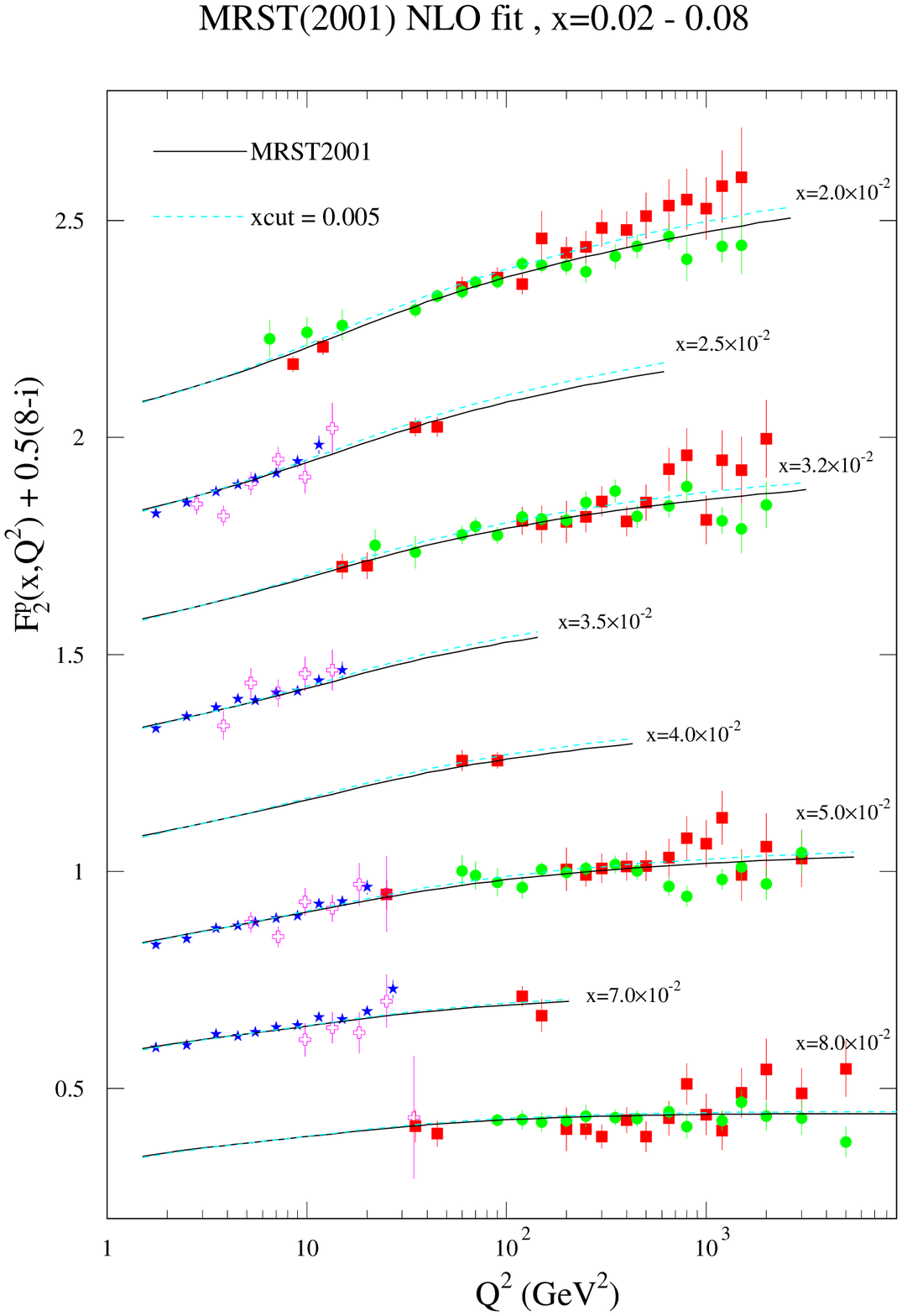}
\includegraphics[width=7cm]{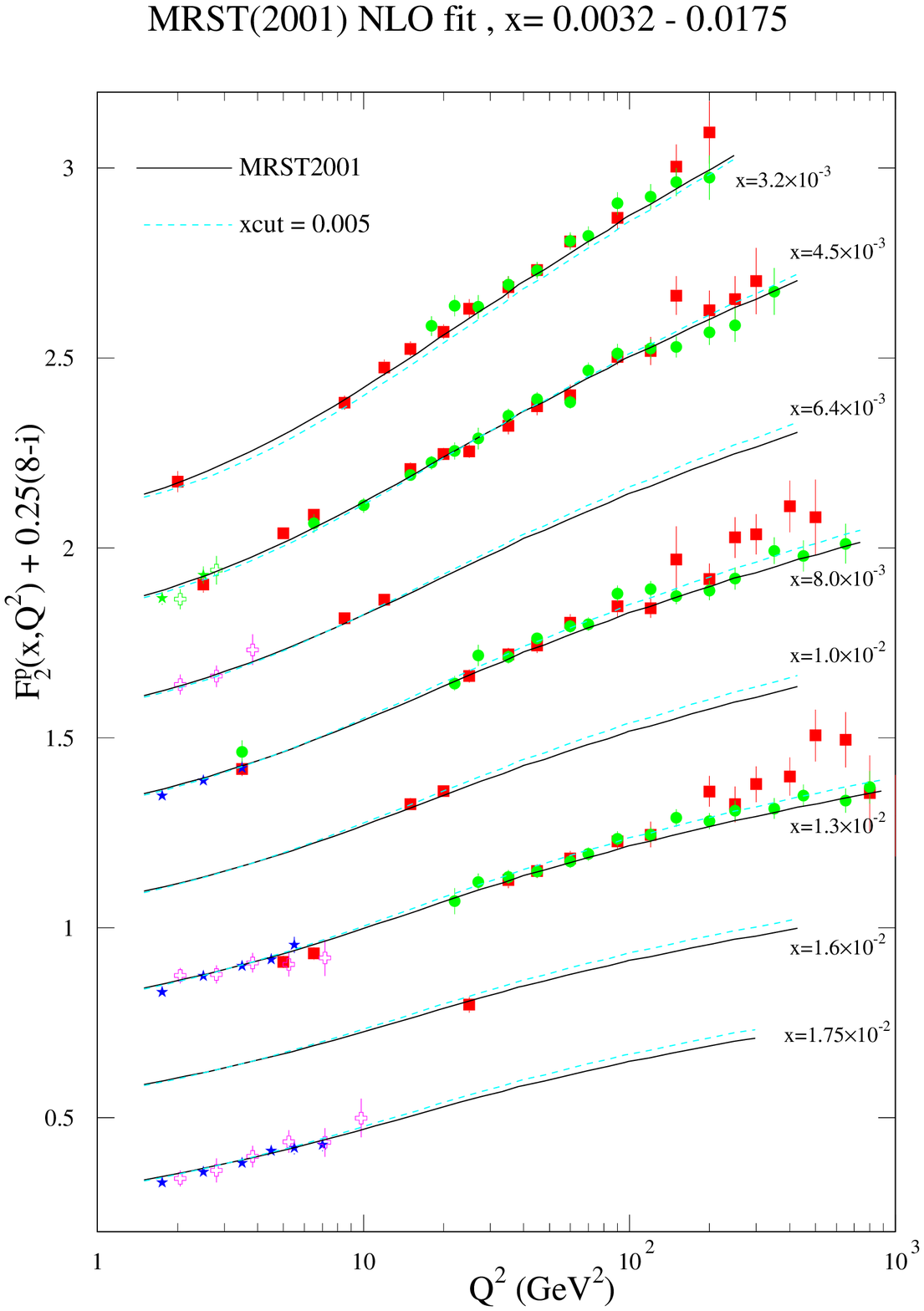}
\vspace{-1.5cm}
\caption{Comparison of MRST(2001) and a fit with $x_{cut}=0.005$.}
\label{six}
\end{figure}

For $W^2_{cut}$ raising from $12.5 \GeV^2$ 
has no effect. Raising $Q^2_{cut}$ from $2\GeV^2$ there is a  
slow continuous improvement for higher $Q^2$ up to 
$> 12 \GeV^2$, suggesting higher order corrections may be important.
The small $x$ gluon decreases slightly as does $\alpha_S(M_Z^2)$
as $Q^2_{cut}$ is raised. 
The predictions for most quantities remain quite stable.  
Raising $x_{cut}$ from $0$ to $0.005$ leads to
continuous improvement - $\Delta \chi^2 =51$ for the data surviving the cut.
The improvement in the fit to structure function data is shown in fig. 6,
and the fit to Tevatron jet data also improves.  
For $x_{cut}=0.005$ there is much reduced tension between different 
data sets. The small $x$ gluon (outside the range of the fit) decreases 
significantly, allowing it to increase for higher $x$, facilitating 
the improved fit. $\alpha_S(M_Z^2)$ falls slightly to $0.118$.
This result suggests that higher order corrections with large 
$\ln(1/x)$ terms could be significant below $x=0.005$. 
With $x_{cut}=0.005$ predictions for Tevatron cross-sections are still 
possible and there is a large change compared to the default fit, as seen 
in fig. 7. The new prediction is well outside the limit set by 
experimental errors, suggesting that the theory error may easily be dominant
for these quantities.  

\begin{figure}[htbp]
\begin{center}
\includegraphics[width=6.7cm]{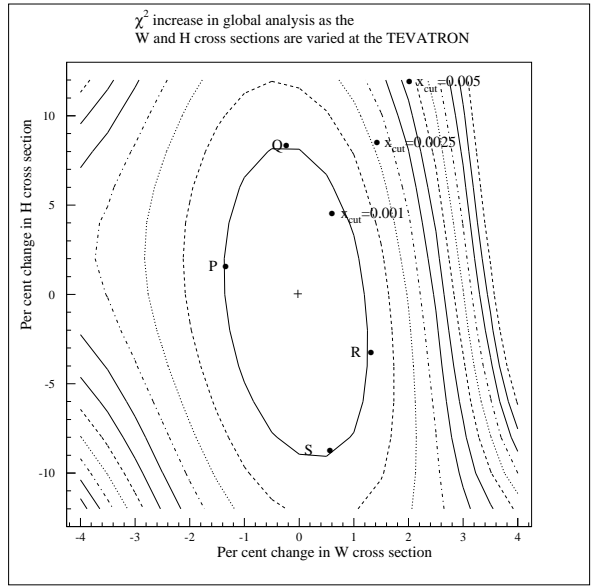}
\end{center}
\vspace{-2.6cm}
\caption{$\Delta \chi^2$-plot for $W$ and Higgs production at the 
Tevatron with $\alpha_S$ free, along with predictions for fits 
with different $x_{cut}$.}
\label{seven}
\end{figure}

\section{Conclusions}

One can perform global fits to data
over a wide range of parameter space determining the partons very
precisely. The fit quality is generally good, but there are some slight 
worries. There are various ways of looking at the uncertainties
on partons due to errors on the data. Although there has been much progress 
recently, there is no universally preferred approach, each having 
strengths and weaknesses. 
The errors on partons and related quantities from this source are 
rather small, i.e. $\sim 1-5 \%$. 

However, the uncertainties from input assumptions 
e.g. cuts on 
data, parameterizations {\it etc}., are comparable and possibly larger.
Also, the errors from higher orders corrections are potentially large,
particularly in some regions of parameter space,
and due to correlations between partons in different regions of phase space 
these feed into all regions (e.g. the small $x$
gluon influences large $x$ gluon).
For some/many processes theory is probably the dominant source of 
uncertainty at present. Systematic study of assumption/theory errors
is needed as well as studies of uncertainties due to errors. 
This is much harder, and is just beginning.      
 
\subsection*{Acknowledgements}

I would like to thank the workshop organizers L.Lyons, M. Whalley and
W.J. Stirling for inviting to present this talk.

\end{document}